\documentclass{nature}
\usepackage{lineno}
\usepackage{float}
\usepackage{graphicx,color}
\usepackage{dcolumn}
\usepackage{amsmath}
\usepackage{epsfig}
\usepackage{amssymb}
\usepackage{amsfonts}

\title{Two-Gap Superconductivity in CaFe$_{0.88}$Co$_{0.12}$AsF Revealed by Temperature Dependence of the Lower Critical Field $H_{c1}^c(T)$}

\author{Teng Wang$^{1,2,3}$, Yonghui Ma$^{1,2,4}$, Wei Li$^{5,6}$, Jianan Chu$^{1,2,4}$, Lingling Wang$^{1}$, Jiaxin Feng$^{1,2,4}$, Hong Xiao$^{7}$, Zhuojun Li$^{1,2}$, Tao Hu$^{1,2}$, Xiaosong Liu$^{1,2,3}$ Gang Mu$^{1,2,*}$}

\begin{document}
\maketitle

\begin{abstract}
Gap symmetry and structure are crucial issues in understanding the superconducting
mechanism of unconventional superconductors. Here we report an in-depth investigation on the out-of-plane lower critical field $H_{c1}^{c}$ of fluorine-based 1111 system superconductor CaFe$_{0.88}$Co$_{0.12}$AsF with $T_c$ = 21 K.
A pronounced two-gap feature is revealed by the kink in the temperature dependent $H_{c1}^c(T)$ curve. The magnitudes of the two gaps are determined to be $\Delta_1$ = 0.86 meV and $\Delta_2$ = 4.48 meV, which account for
74\% and 26\% of the total superfluid density respectively. Our results suggest that the local antiferromagnetic exchange pairing picture is favored compared to the Fermi surface nesting scenario.
\end{abstract}

\footnotetext[1]{State Key Laboratory of Functional Materials for
Informatics, Shanghai Institute of Microsystem and Information
Technology, Chinese Academy of Sciences, Shanghai 200050, China.
$^2$Center for Excellence in Superconducting Electronics (CENSE), Chinese Academy of Sciences, Shanghai 200050, China.
$^3$School of Physical Science and Technology, ShanghaiTech University, Shanghai 201210, China.
$^4$University of Chinese Academy of Sciences, Beijing 100049,
China.
$^5$State Key Laboratory of Surface Physics and Department of Physics, Fudan University, Shanghai 200433, China.
$^6$Collaborative Innovation Center of Advanced Microstructures, Nanjing 210093, China.
$^7$Center for High Pressure Science and Technology Advanced Research, Beijing 100094, China.
Correspondence and requests for materials should be addressed
to G.M. (email: mugang@mail.sim.ac.cn).}

\clearpage

\section*{Introduction}
Superconducting (SC) mechanism is the central issue in the study of unconventional superconductors. Since the discovery of Fe-based superconductors (FeSCs)~\cite{LaFeAsO}, many efforts have been made on this problem~\cite{Mazin2011}.
At the early stage, itinerant mechanism based on the weak correlation was accepted widely and the Fermi surface (FS) nesting (abbreviate as nesting scenario) was believed to be very crucial for the
superconductivity~\cite{Mazin2008,Shoucheng2008}. Later on,
this scenario was challenged by other studies~\cite{TXiang2008,Yildirim2008,QMSi2008,LiFeAs2010}, especially by the discovery of K$_x$Fe$_{2-y}$Se$_2$ system without hole type Fermi surface near the $\Gamma$
point~\cite{XLChen2010,DLFeng2011,HDing2011,XJZhou2011}. Consequently, the
local antiferromagnetic exchange pairing scenario (abbreviate as local scenario), considering a stronger electron correlation, attracts more and more attentions~\cite{JPHu2008,HDingNP2011,HDing2012,JPHu2012}.
Despite the distinct mechanisms mentioned above, the prospective
physical manifestations may be rather subtle. For example, both of them predicted a sign-changed s-wave (S$\pm$) gap symmetry. However, the Fermi surfaces with a better nesting condition tend
to have a stronger pairing amplitude and larger SC gap in the itinerant mechanism~\cite{Mazin2009614,Graser2009,HDing2008}, while according to the local scenario, a larger SC gap should open on the smaller Fermi surface~\cite{JPHu2008}.
Typically approximations were made in the theoretical models
and a precise comparison to the experimental results is difficult. In the case of 122 system Ba$_{0.6}$K$_{0.4}$Fe$_2$As$_2$, the larger SC gap was found to open on the Fermi surfaces with a smaller size and a better
nesting condition~\cite{HDing2008,XJZhou2008,Takahashi2009}, which couldn't discriminate these two theoretical proposals. Therefore, currently more delicate experiments are required.

Recently clear progresses were made on the single-crystal growth of the fluorine-based 1111 system of FeSCs, CaFeAsF~\cite{Ma2015} and the Co doped counterparts~\cite{Ma2016}, and systematic investigations
have been carried out on this system~\cite{Taichi2018,Xiao2016,Xiao2016-2,CaFeAsFCo-Hc2,Xu2018,Ma2018,Gao2018,Mu2018}.
Especially, it was found that the smaller FS around the $\Gamma$ point (see the $\alpha$ FS in Fig. 4) is much smaller than other FSs around $M$ point and consequently shows a worse nesting condition~\cite{CaFeAsFCo-Hc2,Nekrasov},
as compared with the other larger FSs, which should benefit the identification of the abovementioned itinerant and local mechanisms.

In this paper, we present a detailed investigation on the temperature dependence of the out-of-plane lower critical field $H_{c1}^c (T)$
of the high-quality CaFe$_{0.88}$Co$_{0.12}$AsF single crystals. The lower critical field reflects the information of penetration depth and superfluid density, which has been used to investigate the intrinsic SC
properties of FeSCs~\cite{CRen2008,GHCao2018,PhysRevB174512}.
The data is described by a two-component-superfluid model with two SC gaps, $\Delta_1$ = 0.86 meV and $\Delta_2$ = 4.48 meV.
Considering the weighting factors for the two components, we conclude that the larger gap is most likely opened on the smaller Fermi surface, which has a bad nesting condition with other Fermi surfaces.
Thus our results provide a clear identification and the local antiferromagnetic exchange pairing scenario is favored.

\section*{Results}

The dc magnetic susceptibility $\chi$ for the CaFe$_{0.88}$Co$_{0.12}$AsF sample
was measured under a magnetic field of 10 Oe in zero-field-cooling
and field-cooling modes, which is presented in Fig. 1(a). The $\chi-T$ curve shows a sharp SC transition, which reflects the homogeneity and high quality of our sample. The onset transition temperature $T_c$ is about 21 K.
The absolute value of magnetic susceptibility $\chi$ is over 95\% after the demagnetization was considered, indicating a high superconducting volume fraction.
The isothermal $M-H$ curves for the same sample are shown in Figs. 1(b) and (c). The full magnetization curve shown in Fig. 1(b) is rather symmetric, illustrating a very low surface barrier for the flux lines when entering the sample.
For the data in the low-field region as shown in Fig. 1(c), one can see the evolution from the low-field linear tendency to the crooked behavior with the increase of field. The former represents an ideal
Meissner state and the latter reflects the penetration of field into the interior of the sample.

In order to have a clear impression for the data in low-field region, we show the enlarged view of the isothermal $M-H$ curves in Fig. 2(a). The black dashed line represents linear $M-H$ relation in the very low-field
region, which is a consequence of the Meissner effect. Customarily this dashed line is called the Meissner line. We checked the deviation of the magnetization data from the Meissner line to have a solid determination
for the onset point of the field penetration, i.e., $H_{c1}^c$. Field dependence of such a deviation $\Delta M$ is displayed in Fig. 2(b). Two criteria,  $\Delta M$ = 5 $\times$ 10$^{-5}$ emu and 2.5 $\times$ 10$^{-5}$ emu equivalent to
2 Oe and 1 Oe respectively, are adopted for the determination of $H_{c1}^c$. As revealed by the two dashed lines in Fig. 2(b), obviously the variation of criterion will affect the obtained $H_{c1}^c$ values. Nevertheless,
as shown in Fig. 3, the evolution behavior with temperature is not affected by the criterion. In addition, we found that the temperature dependent tendency from our measurements is also consistent with that
obtained by the magnetic torque experiments~\cite{Xiao2016-2}, as displayed by the green asterisks. So we will focus on the analysis of the normalized values $H_{c1}^c(T)/H_{c1}^c(0)$, which are more solid and reliable.

It is known that typically the FeSCs are in the local limit~\cite{CRen2008}, thus the local London model can be used. According to the local London model, the normalized superfluid density
within the $ab$ plane $\widetilde{\rho}_s^{ab}$ has a close relation with the out-of-plane lower critical field $H_{c1}^c$~\cite{CRen2008,GHCao2018}:
\begin{equation}\begin{split}
\widetilde{\rho}_s^{ab}(T) = \frac{\lambda_{ab}^{2}(0)}{\lambda_{ab}^{2}(T)} = \frac{H_{c1}^c(T)}{H_{c1}^c(0)}.\label{eq:1}
\end{split}\end{equation}
Here $\lambda_{ab}$ is the penetration depth within the $ab$ plane.
Moreover, the Fermi surfaces in the present system are nearly ideal cylinders~\cite{CaFeAsFCo-Hc2,Nekrasov}
and the in-plane Fermi velocity is rather isotropic within the $k_x-k_y$ plane. In this case, $\widetilde{\rho}_s^{ab}$ of the $i$th Fermi surface can be given by~\cite{Carrington2003}
\begin{equation}\begin{split}
\widetilde{\rho}_{i}^{ab}(T) = 1+2\int_{\Delta_i}^{\infty}dE\frac{\partial f(E)}{\partial E}\frac{E}{\sqrt{E^2-\Delta_i^2}},\label{eq:2}
\end{split}\end{equation}
where $f(E)$ is the Fermi function and $\Delta_i$ is the value of the energy gap in the $i$th Fermi surface. The temperature dependence of $\Delta_i$ was calculated based on the simple
weak-coupling BCS model. Evidently, the kink feature around $T_c/2$ in Fig. 3 could not be described by an isotropic single gap model.
In order to simplify the discussion, here we adopt a two-gap model and the total normalized superfluid density can be expressed as
\begin{equation}\begin{split}
\widetilde{\rho}_{s}^{ab}(T) = w_1\widetilde{\rho}_{1}^{ab}(T)+w_2\widetilde{\rho}_{2}^{ab}(T),\label{eq:3}
\end{split}\end{equation}
\begin{equation}\begin{split}
w_i = \oint dS_{F,i} \frac{\vec{v_F^{ab}}\cdot\vec{v_F^{ab}}}{v_F^{ab}}.\label{eq:4}
\end{split}\end{equation}
Here $dS_{F,i}$ indicates an integral over the $i$th Fermi surface and $v_F^{ab}$ is the component of Fermi velocity within the $ab$ plane~\cite{Carrington2003}. By tuning the values of $\Delta_i$ and $w_i$,
a simulating curve well describing the experimental
was obtained, as shown by the blue solid curve in Fig. 3. This consistency between our data and the fitting curve suggests that the two-gap model has grasped key features of this system.
The two dashed lines reveal the contributions from the two components with $\Delta_1$ = 0.86 meV, $w_1$ = 0.74, and  $\Delta_2$ = 4.48 meV, $w_2$ = 0.26.

\section*{Discussion}

Investigating the weighting factor $w_{i}$ allows us to seek out the locations of the different superfluid components on the FSs. For the roughly isotropic FSs and isotropic $v_F^{ab}$,
$w_{i}$ is only determined by the $v_F^{ab}$ value and the size of the $i$th FS. Although the detailed correlation is diverse, both the nesting and local scenarios imply that the gap value is determined by the shape and size
of the FSs~\cite{Mazin2009614,Graser2009,HDing2008,JPHu2008}, which can be derived from the calculated electronic structures. As shown in Fig. 4, roughly the five FSs can be divided into two groups from the
viewpoint of FS shape and size: the small $\alpha$ FS and the large ones ($\beta/\gamma/\delta$/$\eta$) with similar sizes.
Thus we only need to simply discuss and compare the two groups. By checking the energy dispersion of the calculated band structure, we estimated that the in-plane $v_F^{ab}$ on the $\alpha$ FS is 1.25$\sim$2 times of that on
the large ones ($\beta/\gamma/\delta$/$\eta$). As for the FS size, however, $\alpha$ FS is only 1/4 of the latter. Moreover, the number of the larger FSs is four, while only one small FS is present.
Considering the above factors, the weighting factor of the $\alpha$ FS should be clearly smaller than that of the others. Consequently we ascribe the $w_2$ and $\Delta_2$ to superfluid on the $\alpha$ FS.

The $\alpha$ FS, which has a rather bad nesting condition with other FSs, carried the superfluid with a larger gap. Evidently, this is inconsistent with the nesting scenario for the pairing mechanism.
Based on the local antiferromagnetic exchange pairing, which considered the local antiferromagnetic exchange of nearest neighboring and next nearest neighbor irons, a simple gap function was proposed~\cite{JPHu2008}:
\begin{equation}\begin{split}
|\Delta(k)| = \Delta_0 |cos k_x cos k_y|.\label{eq:4}
\end{split}\end{equation}
The distribution of the gap value $|\Delta(k)|$ in the brillouin zone was displayed in Fig. 4(a) by the colormap. Obviously the gap value is larger on the smaller $\alpha$ FS, compared with other FSs, which is qualitatively
consistent with our experimental result. So the local scenario is favored for the pairing mechanism of the present system.
Since both the two-gap model and the gap function (eq. 5) are simplified with considerable approximations, we could not carry out a precise and quantitative comparison between the experimental results and the theoretical simulations
at the present stage.

Previously we estimated the value of the in-plane penetration depth $\lambda_{ab}^{ab}(0)$ at 0 K based on the magnetic torque data~\cite{Xiao2016-2}. At that time, only the $\lambda_{ab}^{ab}(T)$ data in
the $T_c/2\leq T\leq T_c$ range were obtained and the kink was not recognized. With the more comprehensive information now, we can update $\lambda_{ab}(0)$ to a more precise value, 260 nm. Based on this value,
we checked the Uemura plot~\cite{Uemura1991} which is a scaling behavior between $T_c$ and $\lambda_{ab}^{-2}$ for the SC systems with a low superfluid density. From Eq. (1),
we have known that $\lambda_{ab}^{-2}$ is proportional to the density of superfluid. As shown in Fig. 4(b), the data of high-$T_c$ cuprates~\cite{Luetkens2008}, FeSCs~\cite{CRen2008,Luetkens2008,Drew2008},
MgB$_2$~\cite{Manzano2002}, and NbSe$_2$~\cite{Fletcher2007} are displayed together.
It is clear that the hole-doped cuprates and the 1111 system of FeSCs reveal a low-superfluid-density feature and follow the linear relation of the Uemura plot. The data point represented by the yellow asterisk is from the
present work and rather consistent with the results of other oxygen-based 1111 systems.

To summarize, we conduct magnetization measurements on CaFe$_{0.88}$Co$_{0.12}$AsF single crystals, and the out-of-plane lower critical field $H_{c1}^c$ is extracted. It is found
that the temperature dependent $H_{c1}^c$ exhibits a pronounced kink around $T_c/2$, which can be described by a two-gap model.
Importantly, the lower superfluid density with a rather large gap is attributed to the small $\alpha$ FS,
from which the local antiferromagnetic exchange pairing mechanism is identified to be a better candidate for understanding the unconventional superconductivity of FeSCs.
Moreover, our data follow the Uemura plot quite well, indicating a low-superfluid-density feature resembling the hole-doped high-$T_c$ cuprates.

\begin{methods}

\subsection{Sample preparation.}

High quality CaFe$_{0.88}$Co$_{0.12}$AsF single crystals were grown using CaAs as
the self-flux~\cite{Ma2015,Ma2016}. The detailed growth conditions and the characterizations of the samples can be seen in our previous reports~\cite{Ma2016}.

\subsection{Magnetization measurements.}
The magnetization measurements were carried out on the magnetic
property measurement system (Quantum Design, MPMS 3). The magnetic fields were applied along the $c$ axis of the single crystal in all the measurements.

\subsection{Band structure calculations.}
The first-principles calculations presented in this work were
performed using the all-electron full potential linear augmented
plane wave plus local orbitals method~\cite{LAPW} as
implemented in the WIEN2K code.~\cite{wien2k} The
exchange-correlation potential was calculated using the generalized
gradient approximation as proposed by Pedrew, Burke, and
Ernzerhof.~\cite{PBE} The calculations for the parent compound were performed using the
experimental crystal structure~\cite{Ma2015}. The band structures for the Co-doped compound were obtained by a slight shift from the results of the parent samples based on a rigid model.

\subsection{Data availability.}
All relevant data are available from the corresponding author.
\end{methods}



\section*{Acknowledgments}
This work is supported by the Youth Innovation Promotion Association of the
Chinese Academy of Sciences (No. 2015187), Natural Science Foundation of China
(No. 11204338 and 11404359), and the ``Strategic Priority Research
Program (B)" of the Chinese Academy of Sciences (No. XDB04040300).

\section*{Additional information}

\subsection{Competing Interests:}
The authors declare no competing financial and non-financial interests.

\section*{Author contributions}
G.M. designed the experiments. T.W. and Y.H.M. synthesized the samples and performed the
measurements. W.L. performed the band structure calculations. G.M. analyzed the data and wrote the paper.
T.W., Y.H.M., W.L., J.N.C., L.L.W., J.X.F., H.X., Z.J.L., T.H., X.S.L., and G.M. discussed the results.
\section*{References}


\clearpage

\section*{Figure captions}

\begin{figure}
\includegraphics[width=17cm]{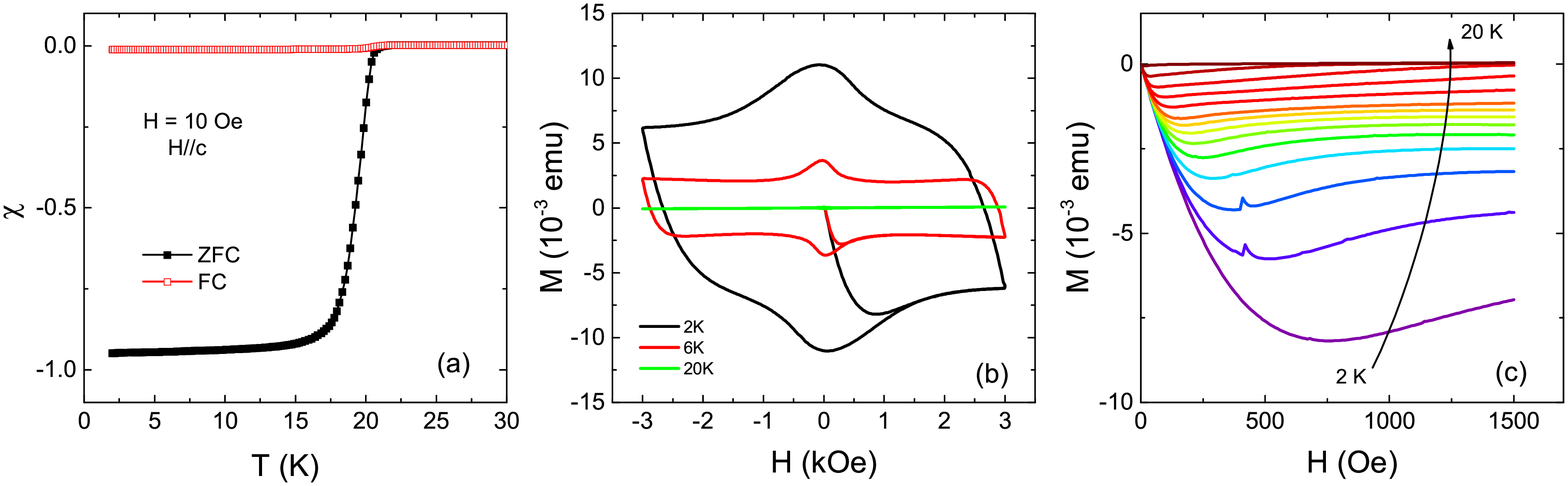}
\caption {Magnetization of the CaFe$_{0.88}$Co$_{0.12}$AsF single crystal. \textbf{a} Temperature dependence of volume susceptibility measured in zero-field-cooled (ZFC) and field-cooled (FC) modes.
\textbf{b} The full magnetization curves at three typical temperatures.
\textbf{c} Field dependence of magnetic moment in the field region below 1500 Oe at different temperatures from 2 K
to 20 K. The intervals are 1 K/step in the range 2-10 K and 2 K/step in the range 10-20 K.} \label{fig1}
\end{figure}

\begin{figure}
\includegraphics[width=7.5cm]{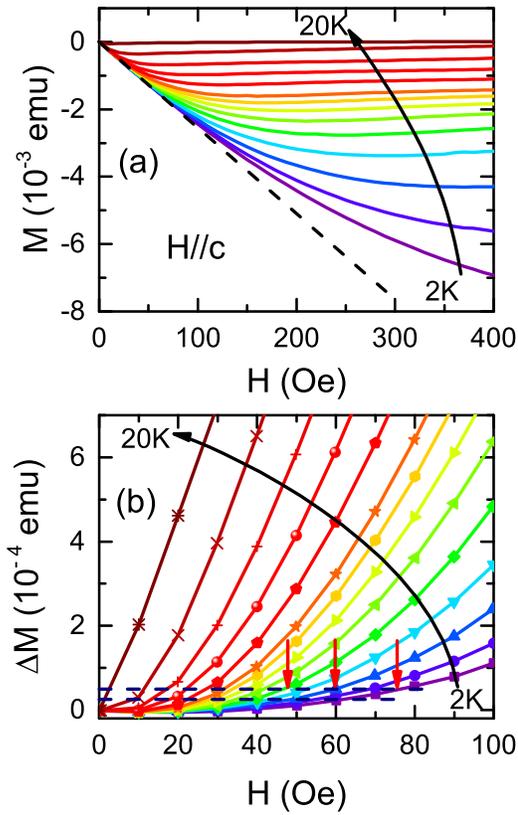}
\caption {Magnetization data in the low-field region.
\textbf{a} Isothermal $M-H$ curves in the field range of 0-400 Oe. The black dashed line shows the linear fit in the low-field region, which is called Meissner line.
\textbf{b} Deviation of magnetization data from the Meissner line. The two dashed lines are two different criteria for determining $H_{c1}^c$. Three red arrows indicate the positions of $H_{c1}^c$ at 2 K, 4 K and 6 K with criteria 1.
The temperature intervals for \textbf{a} and \textbf{b} are the same as those in Figure 1\textbf{c}.}
\label{fig2}
\end{figure}

\begin{figure}
\includegraphics[width=9cm]{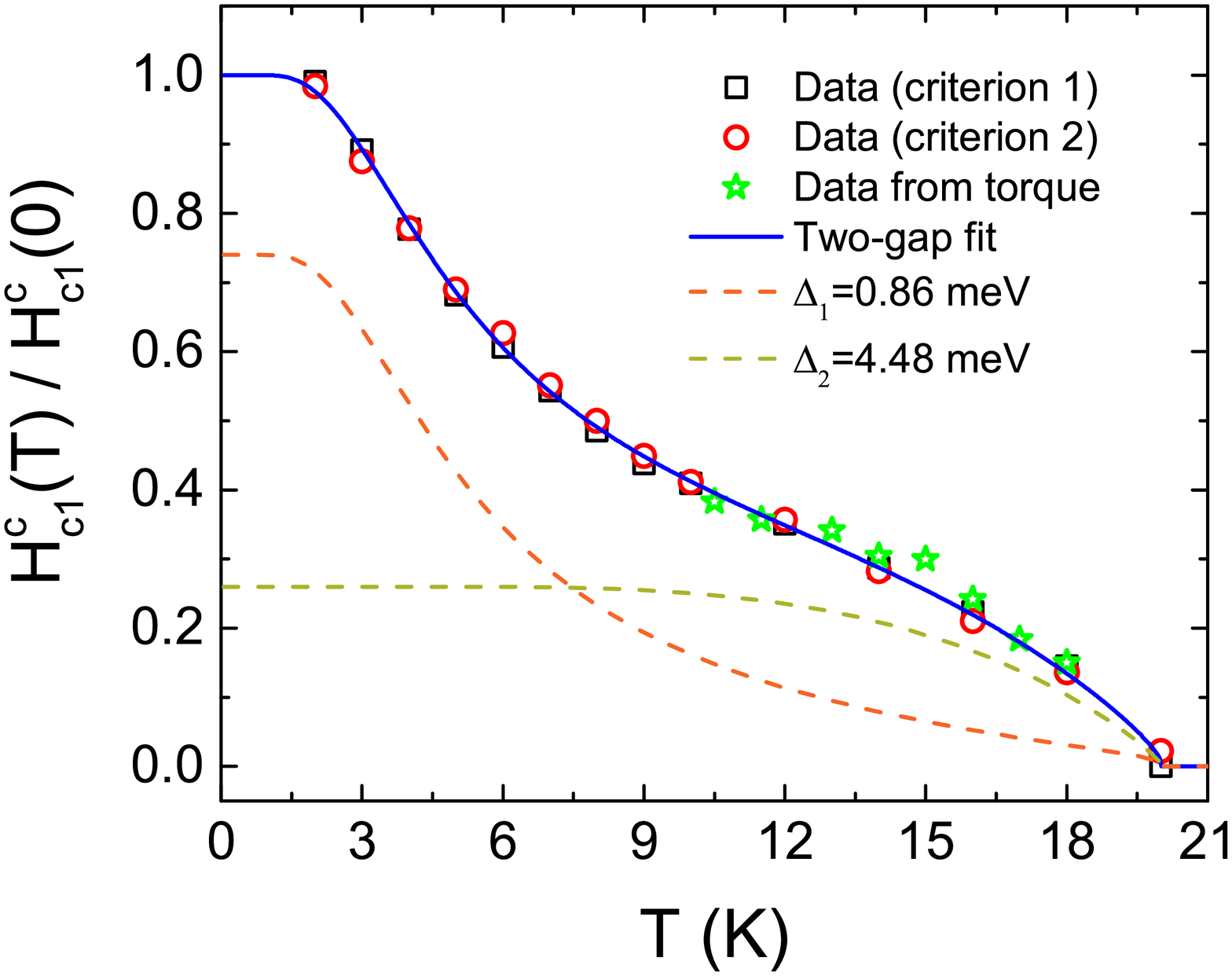}
\caption {The extracted $H_{c1}^c(T)$ (normalized by the zero-temperature value $H_{c1}^c(0)$) as a function of temperature. The data from two criteria are displayed in company with that
from the magnetic torque measurements~\cite{Xiao2016-2}. The solid lines are the fitting curves
using the two-gap model. The contributions of the two components with different gap magnitudes are also shown by the dashed lines.} \label{fig3}
\end{figure}

\begin{figure}
\includegraphics[width=15cm]{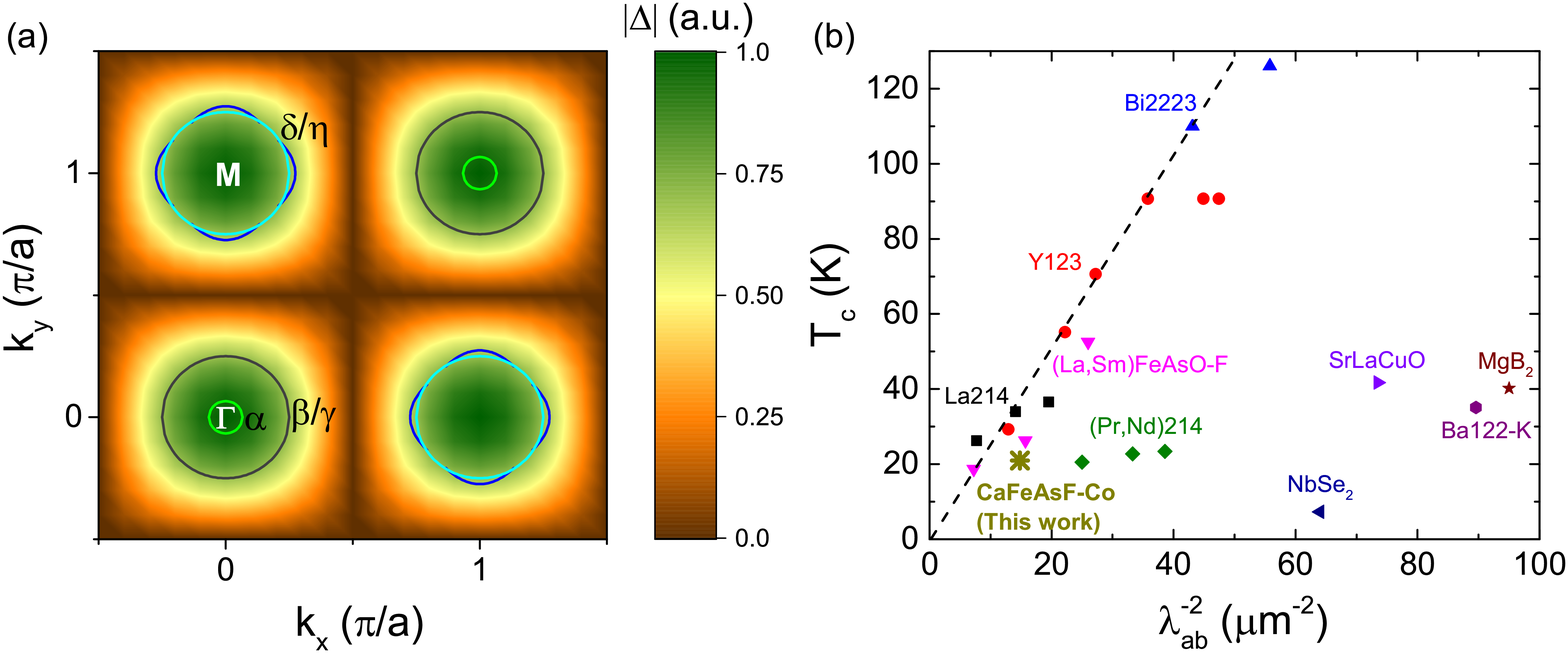}
\caption {Distribution of gap value and the Uemura Plot. \textbf{a} Theoretical SC gap value $|\Delta(k)| = \Delta_0 |cos k_x cos k_y|$ as a function of the
two-dimensional wave vector. The schematic five Fermi surfaces are also shown.
\textbf{b} The correlations between $T_c$ and $\lambda_{ab}^{-2}$, which is proportional to the superfluid density. Points for the cuprates (including La$_{2-x}$Sr$_x$CuO$_{4+\delta}$ (La214), YBa$_2$Cu$_3$O$_{7+\delta}$ (Y123),
Bi$_2$Sr$_2$Ca$_2$Cu$_3$O$_{10+\delta}$ (Bi2223), (Pr,Nd)$_{2-x}$Ce$_x$CuO$_{4+\delta}$ ((Pr,Nd)214), and Sr$_{1-x}$La$_x$CuO$_2$ (SrLaCuO)) and the oxygen-based 1111 system (F-doped LaFeAsO and SmFeAsO ((La,Sm)FeAsO-F)) are taken
from Refs. [38, 39]. The data of K-doped BaFe$_2$As$_2$ (Ba122-K), MgB$_2$, and NbSe$_2$ are taken from
Refs. [33, 40, 41], respectively. The result of CaFe$_{0.88}$Co$_{0.12}$AsF (CaFeAsF-Co) is from the present work.} \label{fig4}
\end{figure}

\end{document}